\documentclass[journal,comsoc]{IEEEtran}

\usepackage[T1]{fontenc}

\ifCLASSINFOpdf
\else
\fi

\usepackage{amsmath}
\usepackage{comment} %
\usepackage{cite}
\usepackage{graphicx}
\usepackage{epsfig,graphics,subfigure,psfrag,amsmath,amssymb}
\usepackage{booktabs}
\usepackage{amsfonts}
\usepackage{epstopdf}
\usepackage{amssymb}
\usepackage{array}
\usepackage{amsmath}
\usepackage{makecell}
\usepackage{multirow}
\usepackage[table,xcdraw]{xcolor}
\usepackage{colortbl}
\usepackage{tabularx}
\usepackage{dblfloatfix}
\usepackage{algorithm}
\usepackage{algorithmic}

\begin{document}
%
\title{Pinching-Antenna Enabled Multicell Wireless Systems}

\author{Yunshu Chen, Qing Xue, \textit{Senior Member, IEEE}, Meng Hua, \textit{Senior Member, IEEE}, Bingpeng Zhou, Shaodan Ma, \textit{Senior Member, IEEE}

\thanks{Manuscript received XXX XX, 2026; revised XXX XX, 202X. This work was supported in part by the National Natural Science Foundation of China under Grant U23A20279.}

\thanks{Yunshu Chen and Qing Xue are with the School of Communications and Information Engineering, Chongqing University of Posts and Telecommunications, Chongqing 400065, China (e-mail: S240101060@stu.cqupt.edu.cn; xueq@cqupt.edu.cn).}

\thanks{Meng Hua is with the Department of Electrical and Electronic Engineering, Imperial College London, SW7 2AZ London, U.K. (e-mail: m.hua@imperial.ac.uk).}

\thanks{Bingpeng Zhou is with the School of Electronics and Communication Engineering, Sun Yat-sen University, Shenzhen 518000, China (e-mail: zhoubp3@mail.sysu.edu.cn).}

\thanks{Shaodan Ma is with the State Key Laboratory of Internet of Things for Smart City and the Department of Electrical and Computer Engineering, University of Macau, Macao SAR, China (e-mail: shaodanma@um.edu.mo).}

}

\maketitle

\begin{abstract}
Pinching antenna (PA) systems have recently emerged as a promising flexible-antenna technology, which can reconstruct the wireless propagation environment by dynamically adjusting the positions of pinching elements along dielectric waveguides, thereby providing new spatial degrees of freedom (DoFs) for enhancing wireless system performance. This paper investigates a multi-waveguide PA-based multi-cell communication system, focusing on the joint optimization of precoding matrices, waveguide power allocation, and antenna placement to maximize the weighted sum rate (WSR).
In multi-cell scenarios, inter-cell interference typically leads to a highly coupled and nonconvex WSR maximization problem. To address this challenge, an efficient alternating optimization framework is adopted to optimize each variable in an iterative way. Specifically, fractional programming is first employed to reformulate the original problem by introducing auxiliary variables that decouple the signal and interference terms. Based on this reformulation, block coordinate descent is then applied to optimize the precoding matrices and power allocation, leading to closed-form or semi-closed-form updates. For the high-dimensional and nonconvex PA placement problem, particle swarm optimization (PSO) is utilized to perform an efficient search and improve scalability.
Numerical results demonstrate that, under various system configurations, the proposed scheme significantly outperforms baseline methods, including average power allocation, fixed antenna placement, conventional multiple-input multiple-output (MIMO), and massive MIMO. These results highlight the strong potential of PA systems for large-scale multi-cell wireless communications.
\end{abstract}


\IEEEpeerreviewmaketitle

\section{Introduction}

With the development of sixth-generation (6G) wireless communication technologies, future networks are expected to support higher data rates, larger-scale connectivity, and more complex communication scenarios. As one of the key technologies in wireless communications, multiple-input multiple-output (MIMO) systems can significantly enhance system capacity through spatial multiplexing gains. However, their performance is largely constrained by the fixed antenna architecture, making it difficult to flexibly adapt to dynamic user distributions and varying propagation environments.
To overcome these limitations, a new class of flexible-antenna technologies has emerged in recent years, including reconfigurable intelligent surfaces (RIS)\cite{2,38}, fluid antennas\cite{3,37}, and movable antennas\cite{4,36}. These technologies reshape the wireless channel by either modifying the electromagnetic propagation environment or finely adjusting antenna positions in a small region, allowing the system to dynamically reconfigure its array response. This, in turn, facilitates more precise beamforming and enhances overall system performance.

Despite their advantages, existing flexible antenna techniques still suffer from inherent limitations. For instance, RIS typically incurs double path loss, while fluid and movable antennas are confined to spatial adjustments within a few wavelengths. These constraints restrict their ability to mitigate large-scale path loss and establish robust line-of-sight (LoS) links\cite{27}, especially in wide-area communication scenarios.
To address these issues, pinching-antenna (PA) systems have emerged as a promising solution with great potential for broad applications\cite{5,10}. This concept was first proposed by NTT DOCOMO, and its feasibility in wireless communications has been experimentally validated\cite{6}.
PA systems utilize dielectric waveguides as the signal transmission medium. By loading small controllable dielectric particles, known as ``pinches'' onto the waveguide surface, localized radiation of electromagnetic waves can be induced at specific positions, thereby forming equivalent radiating antennas\cite{7}. By adjusting the positions of these pinches, both the radiation locations and the phase distribution of the signals can be flexibly controlled, enabling beamforming-like effects.

Compared to existing flexible-antenna technologies, PA systems exhibit several significant advantages. First, PAs can be flexibly deployed along dielectric waveguides over a large spatial range, reaching meter-scale dimensions or beyond. This far surpasses the wavelength-scale adjustment capability of fluid and movable antennas, thereby significantly enhancing the spatial degrees of freedom (DoFs) of the system.
Second, PA systems establish LoS links through direct radiation\cite{8}, avoiding the inherent double path loss issue in RIS-based systems, and thus achieving superior link quality in large-scale propagation environments. Third, PA systems do not require complex or intricate structures; instead, dynamic configuration can be realized through simple dielectric loading, making them structurally simple, cost-effective, and easy to deploy\cite{9}.
More importantly, PA systems not only enable the adjustment of beamforming weights but also allow the joint optimization of antenna positions and power allocation, thereby providing greater design flexibility in the spatial domain.

Benefiting from the aforementioned properties, PA systems can be regarded as a promising enabling technology for next generation wireless communications. They are capable of reshaping the wireless propagation environment with greater precision, thereby, enabling more effective signal enhancement and interference suppression in complex multi-user scenarios\cite{34,35}.

\subsection{Prior Works}

Motivated by the various advantages offered by PA technology, several recent studies have explored its potential. Specifically, \cite{11} investigated the impact of the number of PAs and antenna spacing on the achievable array gain. In\cite{12} and\cite{13}, the authors further considered practical deployment scenarios by establishing analytical frameworks for discretely deployed PA configurations, demonstrating that activating an appropriate number of PAs can significantly enhance system performance.
Building on these insights, existing studies have primarily focused on optimizing PA positions. In particular, \cite{14} studied the rate maximization problem for a downlink single-user PA system, revealing that PA positions jointly affect free-space path loss, free-space phase shifts, and in-waveguide phase shifts. In \cite{15}, the authors extended PA position optimization to multi-user multicast downlink scenarios and proposed a cross-entropy-based optimization framework. In addition, \cite{16} addressed the discrete PA position optimization problem in uplink multi-user multicast scenarios, demonstrating that antenna placement also plays a critical role at the receiver side.

With the advancement of PA-related research, jointly optimizing beamforming \cite{17,18,19,20,23} and power allocation \cite{21,22,24} together with PA placement has attracted increasing attention.
In \cite{17}, the authors jointly optimized transmit beamforming and PA positions, demonstrating that optimizing spatial configuration alone is insufficient to fully exploit system performance. Similarly, \cite{18} established a joint optimization framework for precoding matrices and PA positions, and verified through both uplink and downlink transmissions that the joint design significantly improves system performance and outperforms conventional antenna systems.
Furthermore, \cite{19,20,23} addressed the joint optimization problem of beamforming and antenna placement using different approaches, including element-wise optimization frameworks, gradient meta-learning, and weighted minimum mean square error methods, achieving a favorable trade-off between computational complexity and optimization performance.

Existing studies on power allocation in PA systems are mainly conducted at the user level, improving power utilization by efficiently distributing base station (BS) power among users. In \cite{21}, the authors considered a non-orthogonal multiple access (NOMA)-assisted PA system and jointly optimized PA placement and user power allocation coefficients, improving the overall system rate while satisfying the quality-of-service requirement of the primary user. In \cite{22}, the authors further investigated power control strategies in uplink NOMA systems, demonstrating the critical role of inter-user power allocation in multi-user access scenarios.
Moreover, in wireless powered mobile edge computing systems, the joint optimization of device transmit power, PA placement, and other resource allocation variables was studied in \cite{24}. In particular, \cite{25} and \cite{26} focused on transmit power minimization problems. In \cite{25}, based on the PA power radiation model, the authors jointly optimized beamforming and the number of active antennas, further revealing the coupling between the signal domain and spatial domain. In \cite{26}, beamforming and user power allocation in NOMA-assisted PA systems were jointly considered.
Overall, it can be observed that the focus of current research has gradually evolved from single-variable position optimization to multi-variable joint design involving PA placement, beamforming, and power allocation.
It is worth noting that\cite{41} investigated a simplified multi-cell PA system, where each BS is connected to a single waveguide. By leveraging stochastic geometry, a closed-form expression for the outage probability was derived, which verifies the feasibility and highlights the performance gains of PA systems in multi-cell scenarios.
\subsection{Motivations and Contributions}

Although existing works have explored the potential of PA systems in various system models and communication scenarios, several key challenges remain unresolved. On the one hand, most prior studies focus on single-cell or idealized settings, lacking a systematic investigation of the inter-cell interference and joint resource optimization problems in multi-cell scenarios. On the other hand, when the number of PAs is relatively small, one-dimensional grid search or Gauss-Seidel-based optimization methods can be effectively applied to optimize PA placement. However, in multi-cell systems with a large number of antennas, the dimensionality of the search space increases dramatically, leading to exponentially growing computational complexity, which renders conventional methods impractical.

Motivated by these limitations and inspired by prior works on precoding design in multi-cell systems \cite{29,30,31} as well as the global search capability of heuristic algorithms, we propose to incorporate fractional programming (FP)\cite{33,39} and particle swarm optimization (PSO)\cite{32,40} to address the weighted sum rate (WSR) maximization problem in multi-waveguide PA-based multi-cell communication systems. Specifically, we develop an efficient alternating optimization (AO) framework to jointly optimize the precoding matrices, waveguide power allocation, and PA placement. The main contributions of this paper are summarized as follows:

\begin{itemize}
\item We develop a multi-waveguide PA-based multi-cell multi-user communication model that explicitly accounts for inter-cell interference and waveguide-based channel characteristics, based on which a WSR maximization problem is formulated. Unlike most existing works that focus on user-level power allocation, this paper further considers power allocation across individual PAs. The proposed model captures the coupling among PA placement, precoding matrices, and waveguide power allocation, thereby providing a more accurate reflection of practical system performance.
\item To tackle the resulting nonconvex optimization problem, we propose an equivalent reformulation using FP. By applying the Lagrangian dual transform and quadratic transform, the original multi-ratio WSR maximization problem is converted into a sequence of tractable subproblems, enabling effective decoupling between the signal and interference terms. Building upon this reformulation, FP is further combined with block coordinate descent (BCD) to jointly optimize the precoding matrices and PA power allocation variables. The proposed approach yields closed-form or semi-closed-form updates, significantly reducing computational complexity while ensuring convergence.
\item To address the high-dimensional and nonconvex PA placement optimization subproblem, PSO is employed to perform a efficient search, with constraints handled via penalty functions. Compared with conventional one-dimensional grid search methods, the proposed approach exhibits better scalability for large-scale PA systems.
\item We provide numerical results to validate the effectiveness of the proposed method and compare it with multiple baseline schemes. The results show that: i) compared to average power allocation and fixed PA placement schemes, the joint optimization of waveguide power allocation and placement significantly improves the WSR; ii) under all considered system configurations, the proposed multi-waveguide PA system consistently outperforms both massive MIMO and conventional MIMO systems, demonstrating the great potential of PA technology for large-scale next-generation wireless communication scenarios.
\end{itemize}

\subsection{Organization and Notations}

The remainder of this paper is organized as follows. Section II introduces the PA-based multi-cell communication model and formulates the WSR maximization problem. Section III presents the joint optimization design for the precoding matrix, power allocation, and antenna placement. Section IV provides numerical results. Finally, Section V concludes this paper.

\textit{Notation:} Vectors and matrices are represented by bold lower-case and upper-case letters, respectively. The $N \times N$ identity matrix is denoted by $\mathbf{I}_N$. The sets of real and complex numbers are denoted by $\mathbb{R}$ and $\mathbb{C}$, respectively. The expectation operator is denoted by $\mathbb{E}\{\cdot\}$, and $\mathcal{CN}(\eta,\sigma^2)$ denotes the complex Gaussian distribution with mean $\eta$ and variance $\sigma^2$. The real part of a complex number and the trace operator are denoted by $\Re\{\cdot\}$ and $\mathrm{tr}(\cdot)$, respectively. For any matrix $\mathbf{H}$, $\mathbf{H}^\mathsf{T}$, $\mathbf{H}^*$, and $\mathbf{H}^\mathsf{H}$ denote its transpose, conjugate, and conjugate transpose, respectively.

\section{System Model}

We consider PA-based multi-cell communications systems, as shown in Fig. 1.
In each cell, a BS is connected to $Z$ waveguides, each of which activates $L$ PAs.
We assume that there are $K$ cells, each serving the same number of single-antenna users for notational simplicity.
Let $\text{U}_k^m$ denote the user $m$ in cell $k$, where $m \in \mathcal{M}$ and $k \in \mathcal{K}$.
Define $\mathcal{M}=\{1,2,\ldots,M\}$, $\mathcal{K}=\{1,2,\ldots,K\}$, $\mathcal{Z}=\{1,2,\ldots,Z\}$, and $\mathcal{L}=\{1,2,\ldots,L\}$
as the sets of users per cell, cells, waveguides, and PAs on each waveguide, respectively.

\subsection{Transmission Model}

In cell $k$, the transmit symbol vector is defined as
$\mathbf{s}_k = [s_k^1,\ldots,s_k^M]^\mathsf{T} \in \mathbb{C}^{M\times 1}$,
with $\mathbb{E}[\mathbf{s}_k \mathbf{s}_k^\mathsf{H}] = \mathbf{I}$.
Prior to transmission, the information streams are linearly processed using a digital precoding matrix.
The precoding matrix at the BS $k$ is denoted by $\mathbf{W}_k \in \mathbb{C}^{Z \times M}$,
where $\mathbf{w}_{k,m} \in \mathbb{C}^{Z \times 1}$ represents the precoding vector for user $\text{U}_k^m$.
The transmitted signal from the BS $k$ is then given by
\begin{equation}
\mathbf{x}_k = \mathbf{W}_k \mathbf{s}_k .
\end{equation}

A three-dimensional Cartesian coordinate system is adopted.
In each of the $K$ cells, the $M$ users are located on the $x$-$y$ plane and
distributed within a square region of side length $D$, whose center lies on the $x$-axis.
Each cell is equipped with $Z$ waveguides connected to the BS.
All waveguides are parallel to the $x$-$y$ plane, share the same length $D$, and are positioned at an identical height $d$.
The waveguides are mutually parallel and equally spaced,
and the interference among different waveguides is neglected.
The position of user $\text{U}_k^m$ is denoted by
$\boldsymbol{\psi}_k^m = (x_k^m, y_k^m, 0)$,
where $m \in \mathcal{M}$ and $k \in \mathcal{K}$.
Since $L$ PAs are deployed on each waveguide,
the position of the PA $l$ on the waveguide $z$ in cell $k$ is given by
$\boldsymbol{\psi}_{z,l}^{\mathrm{Pin},k} = (x_{z,l}^k, y_{z,l}^k, d)$,
where $z \in \mathcal{Z}$ and $l \in \mathcal{L}$.
As the inter-waveguide spacing and height are fixed,
we focus on the horizontal coordinates of the PAs.
Define the $x$-axis position matrix of PAs in cell $k$ as
$\mathbf{X}_k \in \mathbb{R}^{L \times Z}$, with $\mathbf{X}_k = [\mathbf{x}_1^k, \mathbf{x}_2^k, \ldots, \mathbf{x}_Z^k]$,
where $\mathbf{x}_z^k = [x_{z,1}^k, x_{z,2}^k, \ldots, x_{z,L}^k]^\mathsf{T} \in \mathbb{R}^{L \times 1}$,
and $x_{z,l}^k \in \left[-\frac{D}{2}, \frac{D}{2}\right]$.

\begin{figure}
 \centering
  \includegraphics[width=\columnwidth]{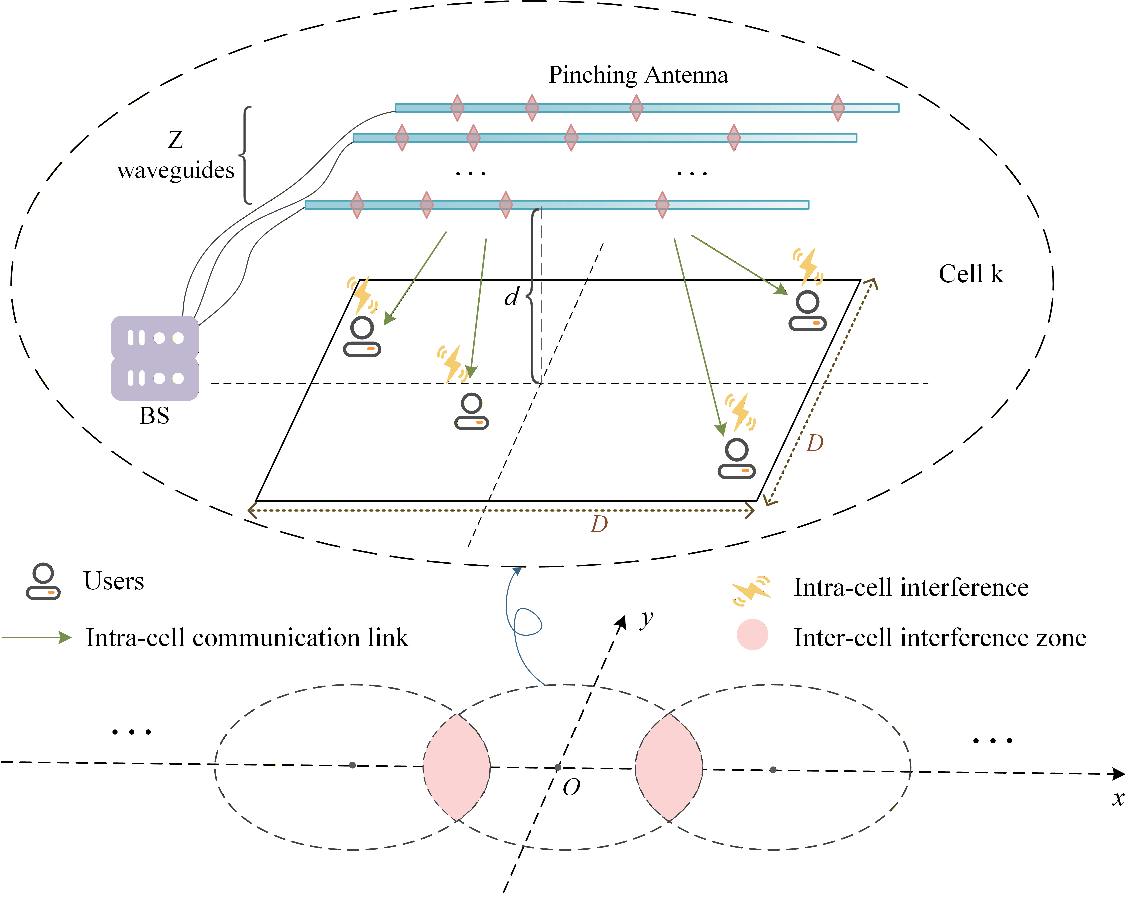}
  \caption{Illustration of the multi-waveguide PA multi-cell communication system.}
  \vspace{-0.7cm}
  \label{fig:1}
  \end{figure}

In general, since the waveguide length is relatively large and the
transmission distance is comparable to the antenna aperture size,
the propagation lies in the near-field region.
Accordingly, we adopt a spherical-wave-based near-field channel model\cite{1}.
In cell $k$, the channel vector from the $L$ PAs on
waveguide $z$ to user $\text{U}_k^m$ is expressed as
\begin{equation}
\begin{gathered}
\makebox[\linewidth][l]{$\mathbf{h}_{k,z,m} =$} \\[2pt]
\left[
\frac{\eta^{\frac12} e^{-j \frac{2\pi}{\lambda}
\left\|\boldsymbol{\psi}_k^m-\boldsymbol{\psi}_{z,1}^{\mathrm{Pin},k}\right\|}}
{\left\|\boldsymbol{\psi}_k^m-\boldsymbol{\psi}_{z,1}^{\mathrm{Pin},k}\right\|},
\cdots,
\frac{\eta^{\frac12} e^{-j \frac{2\pi}{\lambda}
\left\|\boldsymbol{\psi}_k^m-\boldsymbol{\psi}_{z,L}^{\mathrm{Pin},k}\right\|}}
{\left\|\boldsymbol{\psi}_k^m-\boldsymbol{\psi}_{z,L}^{\mathrm{Pin},k}\right\|}
\right]^{\mathsf{T}} ,
\end{gathered}
\tag{2}
\end{equation}
where $\eta = \frac{c^2}{16\pi^2 f_c^2}$ is a constant,
and $c$, $f_c$, and $\lambda$ denote the speed of light,
carrier frequency, and wavelength, respectively.
For PAs deployed on the same waveguide,
the radiated signals are essentially phase-shifted versions
of the signal injected at the BS feeding point.
Let $\boldsymbol{\psi}_{z,0}^{\mathrm{Pin},k}$
denote the position of the feeding point of the waveguide $z$
in cell $k$.
The waveguide response vector can then be written as
\begin{equation}
\mathbf{e}_{k,z}
=
\frac{
\Big[
e^{-j \frac{2\pi}{\lambda_g}
\left\|\boldsymbol{\psi}_{z,0}^{\mathrm{Pin},k} - \boldsymbol{\psi}_{z,1}^{\mathrm{Pin},k}\right\|},
\ldots,
e^{-j \frac{2\pi}{\lambda_g}
\left\|\boldsymbol{\psi}_{z,0}^{\mathrm{Pin},k} - \boldsymbol{\psi}_{z,L}^{\mathrm{Pin},k}\right\|}
\Big]^\mathsf{T}
}
{\sqrt{L}},
\tag{3}
\end{equation}
where $\lambda_g = \frac{\lambda}{n_{\mathrm{eff}}}$ denotes the
waveguide wavelength and $n_{\mathrm{eff}}$ is the effective refractive index of the dielectric waveguides.
Let $p_{k,z,l}$ denote the power allocated to PA $l$
on waveguide $z$ in cell $k$.
Define $\mathbf{P}_{k,z} \in \mathbb{R}^{L \times L}$
as a diagonal matrix whose $l$-th diagonal element is $\sqrt{p_{k,z,l}}$\cite{28}.
Then, the channel coefficient from BS $k$
to user $\text{U}_k^m$ through waveguide $z$ is given by
\begin{equation}
g_{k,m}^{(z)}
=
\mathbf{h}_{k,z,m}^\mathsf{H}
\mathbf{P}_{k,z}
\mathbf{e}_{k,z}.
\tag{4}
\end{equation}
Stacking the contributions from all $Z$ waveguides,
the overall channel vector from BS $k$
to user $\text{U}_k^m$ is
\begin{equation}
\mathbf{g}_{k,m}
=
\left[
g_{k,m}^{(1)},
\ldots,
g_{k,m}^{(Z)}
\right]^\mathsf{T}
\in \mathbb{C}^{Z \times 1}.
\tag{5}
\end{equation}

Based on the above, the received signal at user $\text{U}_k^m$ can be expressed as
\begin{equation}
y_{k,m}
=
\sum_{j=1}^{K}
\mathbf{g}_{j,m}^{\mathsf{T}}
\mathbf{W}_j
\mathbf{s}_j
+
n_{k,m},
\tag{6}
\end{equation}
where $n_{k,m} \sim \mathcal{CN}(0,\sigma^2)$ denotes
the additive white Gaussian noise at user $\text{U}_k^m$.
By adjusting the PA positions
$\boldsymbol{\psi}_{z,l}^{\mathrm{Pin},k}$,
both the phase shifts and path losses in the channel vector
$\mathbf{g}_{k,m}$ can be controlled,
thereby increasing the spatial DoFs of the system.
The signal-to-interference-plusnoise ratio (SINR) at user $\text{U}_k^m$ is given by
\begin{equation}
\mathrm{SINR}_{k,m}
=
\frac{
\left|
\mathbf{g}_{k,m}^{\mathsf{T}}
\mathbf{w}_{k,m}
\right|^2
}
{
\displaystyle
\sum_{\substack{i\neq m}}^{M}
\left|
\mathbf{g}_{k,m}^{\mathsf{T}}
\mathbf{w}_{k,i}
\right|^2
+
\displaystyle
\sum_{\substack{j\neq k}}^{K}
\sum_{i=1}^{M}
\left|
\mathbf{g}_{j,m}^{\mathsf{T}}
\mathbf{w}_{j,i}
\right|^2
+
\sigma^2
}.
\tag{7}
\end{equation}
Accordingly, the achievable rate of user $\text{U}_k^m$ is expressed as
\begin{equation}
R_{k,m}
=
\log_2
\left(
1+\mathrm{SINR}_{k,m}
\right).
\tag{8}
\end{equation}

\subsection{Problem Formulation}

The objective of this study is to jointly optimize
the digital precoding matrices $\mathbf{W}_k$,
the PA position matrices $\mathbf{X}_k$ along the $x$-axis,
and the waveguide power allocation matrices $\mathbf{P}_{k,z}$.
Subject to the BS transmit power budget
and PA mobility constraints,
we aim to maximize the WSR
of all users in the multi-cell network.
This leads to the following optimization problem, denoted as
\begin{equation}
\mathcal{P}_1:\quad
\max_{\mathbf{W}_k, \mathbf{X}_k, \mathbf{P}_{k,z}}
\sum_{k=1}^{K} \sum_{m=1}^{M}
\alpha_{k,m} R_{k,m}
\tag{9}
\end{equation}

\begin{align}
\text{s.t.}\quad
& \mathrm{tr}(\mathbf{W}_k^{\mathsf{H}}\mathbf{W}_k)
\le P_{\max},
\quad \forall k \in \mathcal{K},
\tag{9a}\\
& p_{k,z,l} \ge 0,
\quad \forall l \in \mathcal{L},
k \in \mathcal{K},
z \in \mathcal{Z},
\tag{9b}\\
& \sum_{z=1}^{Z} \sum_{l=1}^{L}
p_{k,z,l}
\le P_{\max},
\quad \forall k \in \mathcal{K},
\tag{9c}\\
& -\frac{D}{2}
\le x_{z,l}^k
\le \frac{D}{2},
\quad \forall l \in \mathcal{L},
k \in \mathcal{K},
z \in \mathcal{Z},
\tag{9d}\\
& x_{z,l+1}^k - x_{z,l}^k
\ge \Delta,
\quad \forall l=1,\ldots,L-1,
k \in \mathcal{K},
z \in \mathcal{Z},
\tag{9e}
\end{align}
where $\alpha_{k,m}$ denotes the non-negative weighting factor
associated with user $\text{U}_k^m$,
and $P_{\max}$ represents the maximum allowable transmit power
of each BS.
Constraint (9b) ensures that the power allocated to each PA
is non-negative.
Constraint (9c) guarantees that the total power allocated to all PAs
within each cell does not exceed the BS power budget.
Constraint (9d) specifies the feasible movement range
of PAs along the $x$-axis,
while (9e) enforces a minimum spacing $\Delta$
between adjacent PAs to avoid severe mutual coupling effects.
Due to the strong coupling among the precoding matrices,
PA positions, and power allocation variables,
the resulting optimization problem is highly nonconvex
and challenging to solve.

\section{Proposed Algorithm For Solving Problem}

\begin{figure*}[!t]
\vspace{-1mm}
\normalsize

\begin{equation}
\widehat{\mathrm{SINR}}_{k,m}
=
\frac{
\left|
\mathbf{g}_{k,m}^{\mathsf{T}}\mathbf{w}_{k,m}
\right|^2
}{
\displaystyle
\sum_{\substack{i\neq m}}^{M}
\left|
\mathbf{g}_{k,m}^{\mathsf{T}}\mathbf{w}_{k,i}
\right|^2
+
\displaystyle
\sum_{\substack{j\neq k}}^{K}
\sum_{i=1}^{M}
\left|
\mathbf{g}_{j,m}^{\mathsf{T}}\mathbf{w}_{j,i}
\right|^2
+
\frac{\sigma^2}{P_{\max}}
\|\mathbf{W}_k\|_F^2
}.
\tag{10}
\end{equation}

\vspace{0.15cm}
\hrule
\vspace{0.15cm}

\begin{equation}
f_R^w(\mathbf W_k,\boldsymbol{\gamma})
=
\sum_{k=1}^{K}\sum_{m=1}^{M}\alpha_{k,m}
\left(
\log(1+\gamma_{k,m})-\gamma_{k,m}
+
\frac{(1+\gamma_{k,m})
\left|
\mathbf g_{k,m}^{\mathsf{T}}\mathbf w_{k,m}
\right|^2}
{
\displaystyle\sum_{j=1}^{K}\sum_{i=1}^{M}
\left|
\mathbf g_{j,m}^{\mathsf{T}}\mathbf w_{j,i}
\right|^2
+
\frac{\sigma^2}{P_{\max}}
\|\mathbf W_k\|_F^2
}
\right).
\tag{13}
\end{equation}
\hrulefill
\end{figure*}

In this section, to solve problem $\mathcal{P}_1$,
we adopt an AO framework
and develop practical and efficient optimization methods
for each group of variables.
For the precoding matrices,
we employ a hybrid approach combining FP and BCD\cite{18},
referred to as the FP-BCD optimization algorithm.
Specifically, the original objective function
is first reformulated into a variational form
without power constraints.
FP is then applied to transform the variational problem
into a Lagrangian dual function
and to perform a quadratic transform\cite{33}.
Subsequently, BCD is used
to iteratively update each block variable,
yielding a locally optimal solution.
Similarly, for the power allocation optimization,
the FP-BCD framework remains applicable.
For the PA position optimization,
since the problem becomes highly nonconvex
and difficult to handle using gradient-based methods,
the PSO algorithm is adopted
to efficiently search for high-quality solutions.

\subsection{Optimize the Precoding Matrices}

Since the original problem involves the joint optimization of
$\mathbf{W}_k$, $\mathbf{P}_{k,z}$, and $\mathbf{X}_k$,
to better illustrate the optimization procedure,
we first fix $\mathbf{P}_{k,z}$ and $\mathbf{X}_k$
and treat $\mathbf{W}_k$ as the optimization variable.
The objective function of problem (9)
is a typical sum-of-logarithmic-ratio form.
It can be reformulated into a variational form
without explicit power constraints.
This transformation implicitly
embeds the power constraint (9a)
into the objective function,
thereby simplifying the constraint handling.
It can be shown that the optimal solution always lies on the boundary of the power constraint\cite{18}.
Therefore, an equivalent $\widehat{\mathrm{SINR}}_{k,m}$ is constructed
as given in eq.(10).

At this stage, the optimization problem $\mathcal{P}_1$
can be rewritten as
\begin{equation}
\mathcal{P}_2:\quad
\max_{\mathbf{W}_k}
\sum_{k=1}^{K}\sum_{m=1}^{M}
\alpha_{k,m}\,\hat{R}_{k,m}
\tag{11}
\end{equation}

\begin{equation}
\text{s.t.}\ (9\text{b})\text{--}(9\text{e}),
\tag{39a}
\end{equation}
where
\(
\hat{R}_{k,m}=\log_2\!\left(1+\widehat{\mathrm{SINR}}_{k,m}\right).
\)
With the power allocation variables $\mathbf{P}_{k,z}$ and
the PA position variables $\mathbf{X}_k$ fixed,
let $\hat{\mathbf{W}}_k$ denote the optimal solution to $\mathcal{P}_2$.
Then, the optimal solution to $\mathcal{P}_1$,
denoted by $\breve{\mathbf{W}}_k$,
can be obtained through the following normalization
\begin{equation}
\breve{\mathbf{W}}_k
=
\sqrt{
\frac{P_{\max}}
{\mathrm{tr}\!\left(\hat{\mathbf{W}}_k^{\mathsf{H}}\hat{\mathbf{W}}_k\right)}
}\,
\hat{\mathbf{W}}_k .
\tag{12}
\end{equation}

Since the power constraint (9a) has been implicitly embedded
into the objective function,
the main challenge of this subproblem lies in the complicated and nonconvex nature of the objective.
To address this, the FP method can be employed.
The basic idea is to first reformulate the logarithmic sum
into a sum-of-ratios form via the Lagrangian dual transform,
and then apply a quadratic transform to each ratio term.
The resulting dual problem can be efficiently solved
using the BCD algorithm.

Specifically, the Lagrangian dual function of $\mathcal{P}_2$ is expressed in eq.(13).
Here, $\gamma_{k,m}\ge0$ is an introduced auxiliary variable,
and $\boldsymbol{\gamma}$ denotes the set $\{\gamma_{k,m}\}$.
According to the Lagrangian duality,
by solving the dual problem
\begin{equation}
(\mathbf W_k^{\max},\boldsymbol{\gamma}^{\max})
=
\arg\max_{\mathbf W_k,\boldsymbol{\gamma}}
f_R^w(\mathbf W_k,\boldsymbol{\gamma}),
\tag{14}
\end{equation}
the solution of $\mathcal{P}_2$ can be obtained indirectly,
as both problems share the same optimal objective value.
The dual problem (14) can be decoupled into two marginal optimization subproblems,
which can be efficiently solved using the BCD algorithm.
It is worth noting that $\mathbf W_k^{\max}$ and $\boldsymbol{\gamma}^{\max}$ denotes the optimal solution to the dual problem in (14), whereas $\boldsymbol{\gamma}^\star$ and $\mathbf W_k^\star$ represent the optimal solutions to the corresponding marginal subproblems in each BCD iteration.

With $\mathbf{W}_k$ fixed, the following marginal optimization
subproblem is considered
\begin{equation}
\boldsymbol{\gamma}^{\star}
=
\arg\max_{\boldsymbol{\gamma}}
f_R^{w}(\mathbf{W}_k,\boldsymbol{\gamma}).
\tag{15}
\end{equation}
This subproblem is convex, and its solution can be obtained by setting
the partial derivative of the objective function to zero, i.e.,
\begin{equation}
\gamma_{k,m}^\star
= \frac{\partial f_R^w}{\partial \gamma_{k,m}}
= \widehat{\mathrm{SINR}}_{k,m}.
\tag{16}
\end{equation}

Let $\gamma_{k,m}=\gamma_{k,m}^{*}$ be fixed,
corresponding to the solution of the first marginal optimization
subproblem. Under this condition, the second marginal optimization
subproblem becomes
\begin{equation}
\mathbf{W}_k^{\star}
=
\arg\max_{\mathbf{W}_k}
f_R^{w}(\mathbf{W}_k,\boldsymbol{\gamma}^{\star}).
\tag{17}
\end{equation}
When $\gamma_{k,m}$ is fixed, only the last ratio-sum term in
$f_R^{w}(\mathbf{W}_k,\boldsymbol{\gamma})$
is involved in the optimization with respect to $\mathbf{W}_k$.
Therefore, subproblem (17) can be further simplified as

\begin{equation}
\mathbf{W}_k^{\star}
=
\arg\max_{\mathbf{W}_k}
\sum_{k=1}^{K}\sum_{m=1}^{M}
\frac{
\alpha_{k,m}(1+\gamma_{k,m}^{\star})
\left|
\mathbf{g}_{k,m}^{\mathsf{T}}\mathbf{w}_{k,m}
\right|^2
}{
\displaystyle\sum_{j=1}^{K}\sum_{i=1}^{M}
\left|
\mathbf{g}_{j,m}^{\mathsf{T}}\mathbf{w}_{j,i}
\right|^2
+
\frac{\sigma^2}{P_{\max}}
\|\mathbf{W}_k\|_F^2
}.
\tag{18}
\end{equation}
The above sum-of-ratios problem can be transformed into a quadratic
form via the quadratic transform.
The resulting quadratic dual function can be written as eq.(19), where
\begin{figure*}[!t]
\vspace{-1mm}
\normalsize

\begin{equation}
f_Q^w(\mathbf W_k,\boldsymbol{\xi})
=
\sum_{k=1}^{K}\sum_{m=1}^{M}\alpha_{k,m}
\Bigg(
2\sqrt{1+\gamma_{k,m}^{\star}}
\,\Re\!\left\{
\xi_{k,m}^{*}\mathbf g_{k,m}^{\mathsf{T}}\mathbf w_{k,m}
\right\}
-
|\xi_{k,m}|^2
\Bigg(
\sum_{j=1}^{K}\sum_{i=1}^{M}
|\mathbf g_{j,m}^{\mathsf{T}}\mathbf w_{j,i}|^2
+
\frac{\sigma^2}{P_{\max}}\|\mathbf W_k\|_F^2
\Bigg)
\Bigg).
\tag{19}
\end{equation}

\vspace{0.15cm}
\hrule
\vspace{0.15cm}

\begin{equation}
\overline{\mathrm{SINR}}_{k,m}
=
\frac{
\left|
\displaystyle\sum_{z=1}^{Z}\bar{\mathbf g}_{k,m}^{{(z)}\,\mathsf{T}}\mathbf p_{k,z}
\right|^{2}
}{
\displaystyle
\sum_{\substack{i\neq m}}^{M}
\left|
\sum_{z=1}^{Z}\bar{\mathbf g}_{k,m}^{{(z)}\,\mathsf{T}}\mathbf p_{k,z}
\right|^{2}
+
\displaystyle
\sum_{\substack{j\neq k}}^{K}
\sum_{i=1}^{M}
\left|
\sum_{z=1}^{Z}\bar{\mathbf g}_{j,m}^{{(z)}\,\mathsf{T}}\mathbf p_{j,z}
\right|^{2}
+\sigma^{2}
}.
\tag{29}
\end{equation}
\hrulefill
\end{figure*}
$\xi_{k,m}\in\mathbb{C}$ denotes the introduced auxiliary
variable and $\boldsymbol{\xi}$ represents the set
$\{\xi_{k,m}\}$.
According to the quadratic duality,
the solution of (18) can be obtained by solving the dual problem
\begin{equation}
\max_{\mathbf W_k,\boldsymbol{\xi}}
f_Q^w(\mathbf W_k,\boldsymbol{\xi}).
\tag{20}
\end{equation}
Since the resulting problem remains nonconvex,
similar to the Lagrangian dual problem,
problem (20) can be approximately solved using the BCD algorithm.

First, by fixing $\mathbf{W}_k$, the optimal value of the
auxiliary variable $\boldsymbol{\xi}$ can be obtained from eq.(19) as
\begin{equation}
\xi_{k,m}^{\star}
=
\frac{
P_{\max}\sqrt{1+\gamma_{k,m}^{\star}}\,
\mathbf{g}_{k,m}^{\mathsf{T}}\mathbf{w}_{k,m}
}{
P_{\max}\displaystyle\sum_{j=1}^{K}
\mathrm{tr}\!\left\{
\mathbf{W}_j\mathbf{W}_j^{\mathsf{H}}
\mathbf{g}_{j,m}^{\ast}\mathbf{g}_{j,m}^{\mathsf{T}}
\right\}
+
\sigma^2\,
\mathrm{tr}\!\left\{
\mathbf{W}_k\mathbf{W}_k^{\mathsf{H}}
\right\}
}.
\tag{21}
\end{equation}

With $\xi_{k,m}^{\star}$ obtained, $\mathbf{W}_k$ can be updated
via eq.(19), i.e.,
\begin{equation}
\mathbf{W}_k^{\star}
=
\arg\max_{\mathbf{W}_k}
f_Q^w(\mathbf{W}_k,\boldsymbol{\xi}^{\star}).
\tag{22}
\end{equation}
Let $\mathbf{G}_k=
\left[
\mathbf{g}_{k,1},
\mathbf{g}_{k,2},
\dots,
\mathbf{g}_{k,M}
\right]
\in\mathbb{C}^{Z\times M}$
denote the matrix of channel vectors in cell $k$.
Then problem (22) can be further rewritten as
\begin{equation}
\mathbf{W}_k^{\star}
=
\arg\max_{\mathbf{W}_k}
F_Q^w(\mathbf{W}_k),
\tag{23}
\end{equation}
where $F_Q^w(\mathbf{W}_k)$ is the matrix-form objective function
derived from eq.(19). This function can be expressed as
\begin{flalign}
&F_Q^w(\mathbf{W}_k)=
\sum_{k=1}^{K}
\Big(
2\Re\{\mathrm{tr}\{\mathbf{T}_k^{\mathsf{H}}\mathbf{G}_k^{\mathsf{T}}\mathbf{W}_k\}\}
 & \notag \\
&-\sum_{j=1}^{K}
\mathrm{tr}\{\mathbf{G}_j^{\mathsf{T}}\mathbf{W}_j\mathbf{W}_j^{\mathsf{H}}\mathbf{G}_j^{\ast}\mathbf{J}_j\}
-\frac{\sigma^2\mathrm{tr}\{\mathbf{J}_k\}\mathrm{tr}\{\mathbf{W}_k\mathbf{W}_k^{\mathsf{H}}\}}{P_{\max}}\Big) , &
\tag{24}
\end{flalign}
with
$\mathbf{T}_k=\mathbf{S}_k\mathbf{U}_k\mathbf{A}_k$
and
$\mathbf{J}_k=\mathbf{S}_k\mathbf{A}_k\mathbf{S}_k^{\mathsf{H}}$,
where $\mathbf{A}_k$, $\mathbf{U}_k$ and $\mathbf{S}_k$
are defined as
\begin{align}
\mathbf{A}_k
&= \mathrm{diag}\{\alpha_{k,1},\ldots,\alpha_{k,M}\},
\tag{25}\\
\mathbf{U}_k
&= \mathrm{diag}\!\left\{
\sqrt{1+\gamma_{k,1}^{\star}},
\ldots,
\sqrt{1+\gamma_{k,M}^{\star}}
\right\},
\tag{26}\\
\mathbf{S}_k
&= \mathrm{diag}\{\xi_{k,1}^{\star},\ldots,\xi_{k,M}^{\star}\}.
\tag{27}
\end{align}

When $\mathbf{P}_{k,z}$ and $\mathbf{X}_k$ are fixed,
problem (23) is convex with respect to the precoding matrix
$\mathbf{W}_k$.
Moreover, the part of the objective function
$F_Q^w(\mathbf{W}_k)$ associated with $\mathbf{W}_k$
can be viewed as a standard regularized linear inverse problem.
Therefore, the closed-form solution of $\mathbf{W}_k$
is given by
\begin{equation}
\mathbf{W}_k^{\star}
=
\left(
\mathbf{G}_k^{\ast}\mathbf{J}_k\mathbf{G}_k^{\mathsf{T}}
+
\frac{\sigma^2 \mathrm{tr}\{\mathbf{J}_k\}}
{P_{\max}}
\mathbf{I}_Z
\right)^{-1}
\mathbf{G}_k^{\ast}\mathbf{T}_k ,
\tag{28}
\end{equation}
where $\sigma^2\mathrm{tr}\{\mathbf{J}_k\}/P_{\max}$ serves as a regularization term.

The proposed FP-BCD algorithm  for precoding matrix optimization is summarized in Algorithm 1.
To obtain the optimal solution for $\mathbf{W}_k$ in $\mathcal{P}_2$,
a two-layer iterative optimization framework is adopted.
In the outer iteration,
the auxiliary variable $\gamma_{k,m}$ is updated according to eq.(16).
For fixed $\gamma_{k,m}$,
the auxiliary variable $\xi_{k,m}$ and the precoding matrix $\mathbf{W}_k$
are alternately updated in the inner iteration.
Specifically,
$\xi_{k,m}$ is first updated according to eq.(21),
followed by the closed-form update of $\mathbf{W}_k$ from eq.(28).
The inner loop repeats until convergence.
After the inner iteration converges,
the algorithm returns to the outer loop to update $\gamma_{k,m}$,
and this process continues until overall convergence.
\begin{algorithm}[!t]
\caption{FP-BCD Algorithm}
\begin{algorithmic}[1]

\REQUIRE Weights $\alpha_{k,m}$ for $k\in\mathcal{K},\, m\in\mathcal{M}$.

\STATE Initialize feasible $\mathbf{W}=\{\mathbf{W}_k\}$.

\REPEAT

\STATE Update dual variable $\gamma_{k,m}$ via (16).

\REPEAT

\STATE Update dual variable $\xi_{k,m}$ via (21).

\STATE Compute $\mathbf{T}_k$ and $\mathbf{J}_k$ via (25)--(27).

\FOR{$k=1$ to $K$}

\STATE Update $\mathbf{W}_k$ via (28).

\ENDFOR

\UNTIL convergence

\UNTIL convergence

\end{algorithmic}
\end{algorithm}

In each outer iteration, each auxiliary variable $\gamma_{k,m}$ requires computing an equivalent SINR. This process involves summation over all inter-cell and inter-user interference terms. Therefore, updating all $\gamma_{k,m}$ incurs a complexity of $\mathcal{O}(K^2 M^2 Z)$. In each inner iteration, the update of the auxiliary variable $\xi_{k,m}$ follows a similar procedure, leading to the same complexity of $\mathcal{O}(K^2 M^2 Z)$.
In addition, the complexity of the constructed diagonal matrices $\mathbf{T}_k$ and $\mathbf{J}_k$ is $\mathcal{O}(KM)$, which is relatively small and can be neglected. Finally, updating the precoding matrices $\mathbf{W}_k$ for all cells requires $\mathcal{O}(K(Z^3 + Z^2 M))$ operations.
Therefore, the overall computational complexity of Algorithm 1 can be expressed as
$\mathcal{O}\big(I_{\mathrm{out}}^{(W)}[K^2 M^2 Z + I_{\mathrm{in}}^{(W)}(K^2 M^2 Z + K(Z^3 + Z^2 M))]\big)$,
where $I_{\mathrm{out}}^{(W)}$ and $I_{\mathrm{in}}^{(W)}$ denote the numbers of outer and inner iterations for precoding matrices optimization, respectively.

\subsection{Optimize the Power Allocation Variables}

\begin{figure*}[!t]
\vspace{-1mm}
\normalsize

\begin{equation}
f_R^p(\mathbf P_{k,z},\boldsymbol{\chi})
=
\sum_{k=1}^{K}\sum_{m=1}^{M}\alpha_{k,m}
\left(
\log(1+\chi_{k,m})-\chi_{k,m}
+
\frac{(1+\chi_{k,m})
\left|
\displaystyle\sum_{z=1}^{Z}\bar{\mathbf g}_{k,m}^{z\,\mathsf{T}}\mathbf p_{k,z}
\right|^2}
{
\displaystyle
\sum_{j=1}^{K}\sum_{i=1}^{M}
\left|
\sum_{z=1}^{Z}\bar{\mathbf g}_{j,m}^{z\,\mathsf{T}}\mathbf p_{j,z}
\right|^2
+
\sigma^2
}
\right).
\tag{31}
\end{equation}

\vspace{0.15cm}
\hrule
\vspace{0.15cm}

\begin{equation}
f_Q^p(\mathbf P_{k,z},\boldsymbol{\varepsilon})
=
\sum_{k=1}^{K}\sum_{m=1}^{M}
\left(
2\sqrt{\alpha_{k,m}(1+\chi_{k,m}^{\star})}
\,\Re
\left\{
\varepsilon_{k,m}^{*}
\sum_{z=1}^{Z}
\bar{\mathbf g}_{k,m}^{z\,\mathsf{T}}\mathbf p_{k,z}
\right\}
-
|\varepsilon_{k,m}|^2
\left(
\sum_{j=1}^{K}\sum_{i=1}^{M}
\left|
\sum_{z=1}^{Z}
\bar{\mathbf g}_{j,m}^{z\,\mathsf{T}}\mathbf p_{j,z}
\right|^2
+
\sigma^2
\right)
\right).
\tag{33}
\end{equation}
\hrulefill
\end{figure*}

With $\mathbf W_k$ and $\mathbf X_k$ fixed,
we optimize the power allocation variables $\mathbf P_{k,z}$.
Since the power variables are implicitly embedded in the channel vectors,
an equivalent $\overline{\mathrm{SINR}}_{k,m}$ is constructed based on eq.(7),
as given in eq.(29).
This equivalent SINR formulation explicitly separates the power allocation variables from the channel representation.
Let $\mathbf p_{k,z}\in\mathbb C^{L\times1}$ denote the column vector formed from
the diagonal matrix $\mathbf P_{k,z}$.
$\overline{\mathbf g}_{k,m}^{(z)}\in\mathbb C^{L\times1}$ denotes the equivalent
channel vector from the BS to user $\text{U}_k^m$ through waveguide $z$
in cell $k$.
The $l$-th element of this vector can be written as
$\frac{
\eta^{\frac12}\omega_z
e^{-j\left(
\frac{2\pi}{\lambda}\left\|\boldsymbol{\psi}_k^m-\boldsymbol{\psi}_{z,l}^{\mathrm{Pin},k}\right\|
+
\frac{2\pi}{\lambda_g}\left\|\boldsymbol{\psi}_{z,0}^{\mathrm{Pin},k}-\boldsymbol{\psi}_{z,l}^{\mathrm{Pin},k}\right\|
\right)}
}{
\sqrt{L}\,\left\|\boldsymbol{\psi}_k^m-\boldsymbol{\psi}_{z,l}^{\mathrm{Pin},k}\right\|
}$\cite{28},
where $\omega_z$ denotes the element of $\mathbf{w}_{k,m}\in\mathbb{C}^{Z\times1}$.
The original optimization problem $\mathcal{P}_1$ can then be reformulated as
\begin{equation}
\mathcal{P}_3:\quad
\max_{\mathbf{P}_{k,z}}
\sum_{k=1}^{K}\sum_{m=1}^{M}\alpha_{k,m}\,\overline{R}_{k,m}
\tag{30}
\end{equation}

\begin{equation}
\text{s.t.}\ (9\text{b})\text{--}(9\text{c}),
\tag{30a}
\end{equation}
with
$\overline{R}_{k,m}=\log_2(1+\overline{\mathrm{SINR}}_{k,m})$.
Similar to the optimization of the precoding matrix $\mathbf{W}_k$ in the previous subsection,
the objective function of problem $\mathcal{P}_3$ also takes a sum-of-logarithmic-ratios form.
Hence, the FP-BCD framework can also be applied,
where the optimal solution is obtained via the Lagrangian dual transform and the quadratic transform.

Since the detailed steps of the FP-BCD algorithm have been described in the previous subsection, we briefly summarize the key steps here.
First, the Lagrangian dual function of the objective function in (30) is expressed in eq.(31), where
$\chi_{k,m}$ denotes the introduced auxiliary variable, and
$\boldsymbol{\chi}$ represents the set $\{\chi_{k,m}\}$.
For fixed $\mathbf p_{k,z}$, the optimal $\chi_{k,m}^{\star}$ can be obtained as
\begin{equation}
\chi_{k,m}^{\star}
=
\overline{\mathrm{SINR}}_{k,m}.
\tag{32}
\end{equation}
Next, with $\chi_{k,m}^{\star}$ fixed, the variable $\mathbf p_{k,z}$ is optimized.
In this case, only the last ratio-sum term in
$f_R^p(\mathbf P_{k,z},\boldsymbol{\chi})$ contains the optimization variable
$\mathbf p_{k,z}$. Therefore, by applying the quadratic transform,
$f_R^p(\mathbf P_{k,z},\boldsymbol{\chi})$ can be reformulated as in eq.(33)
where $\varepsilon_{k,m}$ denotes the introduced auxiliary variable and
$\boldsymbol{\varepsilon}$ represents the set $\{\varepsilon_{k,m}\}$.
Similarly, the update equation of $\boldsymbol{\varepsilon}$ can be derived from eq.(33) as
\begin{equation}
\varepsilon_{k,m}^{\star}
=
\frac{
\sqrt{\alpha_{k,m}\left(1+\chi_{k,m}^{\star}\right)}
\displaystyle\sum_{z=1}^{Z}\bar{\mathbf g}_{k,m}^{(z)\,\mathsf{T}}\mathbf p_{k,z}
}{
\displaystyle
\sum_{j=1}^{K}\sum_{i=1}^{M}
\left|
\sum_{z=1}^{Z}\bar{\mathbf g}_{j,m}^{(z)\,\mathsf{T}}\mathbf p_{j,z}
\right|^2
+\sigma^2
}.
\tag{34}
\end{equation}

\begin{algorithm}[!t]
\caption{PSO-based PA placement Algorithm}
\begin{algorithmic}[1]

\STATE \textbf{Initialize:} randomly generate $N$ particles $\{\mathbf r_n\}$ within the feasible search space, and initialize $u=1$, $b_0$, $b_1$, and $b_2$.

\STATE Evaluate the fitness function $f(\mathbf r_n)$ for each particle.

\STATE Set $\boldsymbol{\mathit{pbest}}_n=\mathbf r_n^{0}$, and set $\mathbf{gbest}$ as the particle with the maximum fitness among all $\mathbf{pbest}_n$.

\REPEAT

\FOR{$n=1$ to $N$}

\STATE Update the position $\mathbf r_n^{u}$ and velocity $\mathbf v_n^{u}$ based on (42).

\IF{$f(\mathbf r_n^{u}) \geq f(\boldsymbol{\mathit{pbest}}_n)$}
    \STATE $\boldsymbol{\mathit{pbest}}_n=\mathbf r_n^{u}$.
\ENDIF

\IF{$f(\mathbf r_n^{u}) \geq f(\boldsymbol{\mathit{gbest}})$}
    \STATE $\boldsymbol{\mathit{gbest}}=\mathbf r_n^{u}$.
\ENDIF

\ENDFOR

\STATE Update $u=u+1$.

\UNTIL{convergence}

\end{algorithmic}
\end{algorithm}

Define the vector
$\mathbf p_k=[\mathbf p_{k,1}^\mathsf{T},\ldots,\mathbf p_{k,Z}^\mathsf{T}]^\mathsf{T}\in\mathbb{R}^{ZL}$
and
$\bar{\mathbf g}_{k,m}
=
[\bar{\mathbf g}_{k,m}^{(1)\,\mathsf{T}},\ldots,\bar{\mathbf g}_{k,m}^{(Z)\,\mathsf{T}}]
\in\mathbb{C}^{1\times ZL}$.
Then, the quadratic dual function in eq.(33) with respect to $\mathbf P_{k,z}$
can be rewritten in a standard quadratic form as
\begin{equation}
f_Q^p(\mathbf P_{k,z})
=
2\mathbf b_k^\mathsf{T}\mathbf p_k
-
\mathbf p_k^\mathsf{T}\mathbf E_k\mathbf p_k ,
\tag{35}
\end{equation}
where
\begin{equation}
\mathbf b_k
=
\sum_{m=1}^{M}
\sqrt{\alpha_{k,m}(1+\chi_{k,m}^{\star})}
\,\Re\!\left\{\varepsilon_{k,m}^{\star}\bar{\mathbf g}_{k,m}^{\mathsf{H}}\right\}\in\mathbb{R}^{ZL\times 1},
\tag{36}
\end{equation}

\begin{equation}
\mathbf E_k
=
\sum_{r=1}^{K}\sum_{m=1}^{M}
|\varepsilon_{r,m}^{\star}|^2
\,\Re\!\left\{\bar{\mathbf g}_{r,m}^{\mathsf{H}}\bar{\mathbf g}_{r,m}\right\}\in\mathbb{R}^{ZL\times ZL}.
\tag{37}
\end{equation}
Thus, given $\varepsilon_{k,m}^{\star}$,
the semi-closed-form solution of the power variable $\mathbf p_k$ is obtained as
\begin{equation}
\mathbf p_k^{\star}
=
(\mathbf E_k+\mu_k\mathbf I)^{-1}\mathbf b_k ,
\tag{38}
\end{equation}
where $\mu_k\ge0$ is the total power multiplier introduced by the
power constraint (9c).
Similar to the optimization of $\mathbf W_k$,
an outer-inner iterative procedure is adopted to obtain the optimized
power allocation variables in $\mathcal{P}_3$.

Since the power allocation optimization also adopts the FP-BCD framework, its complexity analysis follows a similar procedure to that of Algorithm 1. In each outer iteration, updating the auxiliary variable $\chi_{k,m}$ incurs a complexity of $\mathcal{O}(K^2 M^2 ZL)$. In each inner iteration, the update of the auxiliary variable $\varepsilon_{k,m}$ has the same complexity of $\mathcal{O}(K^2 M^2 ZL)$.
Then, constructing the matrix $\mathbf{E}_k$ requires $\mathcal{O}(K^2 M (ZL)^2)$ operations, while constructing the vector $\mathbf{b}_k$ has a complexity of $\mathcal{O}(K M ZL)$, which is relatively small and can be neglected. Finally, updating the power vectors $\mathbf{p}_k$ for all cells requires $\mathcal{O}(K (ZL)^3)$ operations.
Therefore, the overall computational complexity of the power allocation subproblem can be expressed as
$\mathcal{O}\big(I_{\mathrm{out}}^{(P)}[K^2 M^2 ZL + I_{\mathrm{in}}^{(P)}(K^2 M^2 ZL + K^2 M (ZL)^2 + K (ZL)^3)]\big)$, where $I_{\mathrm{out}}^{(P)}$ and $I_{\mathrm{in}}^{(P)}$ denote the numbers of outer and inner iterations.

\subsection{Optimize the PA Positions}

Due to the large number of PAs in the multi-waveguide multi-cell system,
the antenna position optimization becomes a high-dimensional
nonconvex subproblem.
As a result, conventional convex optimization methods
are generally ineffective for solving the PA placement problem.
To address this issue, the PSO
algorithm is adopted.
With $\mathbf{W}_k$ and $\mathbf{P}_{k,z}$ fixed,
problem $\mathcal{P}_1$ can be reformulated as
\begin{equation}
\mathcal{P}_4:\quad
\max_{\mathbf{X}_k}
\sum_{k=1}^{K}\sum_{m=1}^{M}\alpha_{k,m}R_{k,m}
\tag{39}
\end{equation}

\begin{equation}
\text{s.t.}\ (9\text{d})\text{--}(9\text{e}).
\tag{39a}
\end{equation}
The PSO algorithm is a low-complexity and easy-to-implement method,
which is capable of performing efficient search
for high-dimensional problems\cite{32}.
To better apply this algorithm,
the objective function in $\mathcal{P}_4$ is simplified,
and the problem is further reformulated into the following
unconstrained form
\begin{equation}
\mathcal{P}_5:\quad
\max_{\mathbf{X}_k}
\sum_{k=1}^{K}\sum_{m=1}^{M}\alpha_{k,m}\mathrm{SINR}_{k,m}
-
\Pi(\mathbf{X}_k)
\tag{40}
\end{equation}
where $\Pi(\mathbf{X}_k)$ is the penalty term associated with
constraint (39a), defined as
\begin{equation}
\Pi(\mathbf{X}_k)
=
\begin{cases}
0, & \text{if } (39\text{a}) \text{ holds},\\
+\infty, & \text{otherwise}.
\end{cases}
\tag{41}
\end{equation}

In the PSO algorithm,
each particle represents a potential solution.
In the proposed PA position optimization problem,
each particle corresponds to a candidate configuration of all PA positions in the system.
By continuously updating its position and velocity,
each particle searches the solution space and gradually approaches the optimal solution.
First, $N$ particles are randomly initialized in the feasible search space,
which is determined by the allowable placement range on each waveguide and the minimum spacing constraints between adjacent PAs,
with their coordinates expressed as
$\mathbf r_n=\{r_{1,n},\ldots,r_{K\times Z\times L,n}\},\ n=1,\ldots,N$.
During the optimization process,
the current position of each particle is substituted into the objective
function in (39) to obtain its fitness value $f(\mathbf r_n)$ based on the achieved WSR.
The personal best position $\boldsymbol{\mathit{pbest}}_n$ is then updated.
By comparing the fitness values of all particles,
the global best position $\boldsymbol{\mathit{gbest}}$ is determined
from the set of personal best positions $\{\boldsymbol{\mathit{pbest}}_n\}$.
At the $u$-th iteration,
the position and velocity of each particle are updated as
\begin{equation}
\begin{aligned}
\mathbf r_n^{u} &= \mathbf r_n^{u-1} + \mathbf v_n^{u},\\
\mathbf v_n^{u} &= b_0 \mathbf v_n^{u-1}
+ b_1 c_1 (\boldsymbol{\mathit{pbest}}_n - \mathbf r_n^{u-1})
+ b_2 c_2 (\boldsymbol{\mathit{gbest}} - \mathbf r_n^{u-1}),
\end{aligned}
\tag{42}
\end{equation}
where $b_0$ denotes the inertia weight, representing the influence of the
previous velocity on the current velocity,
$b_1$ and $b_2$ are the learning coefficients,
and $c_1$ and $c_2$ are random variables uniformly distributed in $[0,1]$.
Here, $\mathbf r_n^{u}$ and $\mathbf v_n^{u}$ denote the position and velocity of the $n$-th particle in the $u$-th iteration, respectively.
The algorithm iterates until the fitness value converges,
after which the optimized PA positions in $\mathcal{P}_4$ are obtained.
The overall PSO algorithm is summarized in Algorithm 2.

The computational complexity  of the PSO algorithm mainly arises from the evaluation of the fitness function and the update of particle positions and velocities. For each particle, the fitness function is given by eq.(40), which requires computing the SINR of all users. Therefore, the complexity of a single fitness evaluation is $\mathcal{O}(K^2 M^2 Z)$.
In addition, updating the position and velocity of each particle involves operations over all PA position variables with dimension $K Z L$, leading to a complexity of $\mathcal{O}(K Z L)$. Therefore, the overall computational complexity of Algorithm 2 is $\mathcal{O}\big(I_{\mathrm{pso}} N (K^2 M^2 Z + K Z L)\big)$, where $I_{\mathrm{pso}}$ denotes the number of PSO iterations.

\begin{algorithm}[t]
\caption{Overall AO Algorithm}
\begin{algorithmic}[1]

\STATE \textbf{Initialize:} Feasible $\mathbf{W}=\{\mathbf{W}_k\}$,
$\mathbf{P}=\{\mathbf{P}_{k,z}\}$ and $\mathbf{X}=\{\mathbf{X}_k\}$,
initialize AO iteration count $i=1$.

\REPEAT

\STATE Fix $\mathbf{P}^{i-1}$, $\mathbf{X}^{i-1}$, update $\mathbf{W}^i$
by solving problem (11) through Algorithm 1.

\STATE Fix $\mathbf{W}^i$, $\mathbf{X}^{i-1}$, update $\mathbf{P}^i$
by solving problem (30) through Algorithm 1.

\STATE Fix $\mathbf{W}^i$, $\mathbf{P}^i$, update $\mathbf{X}^i$
by solving problem (39) through Algorithm 2.

\STATE Set $\mathbf{X}^i=\boldsymbol{\mathit{gbest}}$ obtained from PSO,
and compute $\mathrm{WSR}^i$ using $\mathbf{W}^i$, $\mathbf{P}^i$ and $\mathbf{X}^i$.

\IF{$\mathrm{WSR}^i < \mathrm{WSR}^{i-1}$}
\STATE $\mathbf{W}^i=\mathbf{W}^{i-1}$,
$\mathbf{P}^i=\mathbf{P}^{i-1}$,
$\mathbf{X}^i=\mathbf{X}^{i-1}$.

$\mathrm{WSR}^i=\mathrm{WSR}^{i-1}$.
\ENDIF

\STATE Update $i=i+1$.

\UNTIL{convergence}

\STATE Scale $\mathbf{W}$ as
$\mathbf{W}=\sqrt{\frac{P_{\max}}{\mathrm{tr}(\mathbf{W}^\mathsf{H}\mathbf{W})}}\mathbf{W}$.

\STATE \textbf{return} Precoding matrix $\mathbf{W}$,
Power allocation matrix $\mathbf{P}$ and position matrix $\mathbf{X}$.

\end{algorithmic}
\end{algorithm}

\subsection{Overall Algorithm}

\textit{1) Convergence Analysis:} Based on the above derivations for the precoding matrix $\mathbf{W}_k$,
the power allocation matrix $\mathbf{P}_{k,z}$, and the PA placement
$\mathbf{X}_k$, the original problem $\mathcal{P}_1$ can be solved through an
AO framework, in which these variable sets are
updated in an iterative manner.
The overall AO procedure is summarized in Algorithm 3.
Specifically, in each iteration,
the precoding matrix $\mathbf{W}_k$ and the power allocation matrix
$\mathbf{P}_{k,z}$ are updated using the FP-BCD algorithm, while the PA
placement $\mathbf{X}_k$ is optimized by employing the PSO algorithm.

It is worth noting that the FP-BCD based updates for $\mathbf{W}_k$ and $\mathbf{P}_{k,z}$
guarantee a non-decreasing objective value. However, since the PSO
algorithm is a stochastic search method that does not strictly ensure
the improvement of the objective value in each iteration, the obtained
solution may occasionally lead to a smaller WSR
than that of the previous AO iteration. To address this issue and
ensure the stability of the iterative process, a rollback mechanism
is introduced. Specifically, if the WSR achieved in the current
iteration is lower than that in the previous iteration, the updated
variables are discarded and the previous solution is retained.

\textit{2) Complexity Analysis:} Since the computational complexities of Algorithm 1 and Algorithm 2 have been analyzed previously, the overall complexity of Algorithm 3 can be expressed as $\mathcal{O}\big(I_{\mathrm{AO}}(C_W + C_P + C_X)\big)$, where $C_W$, $C_P$, and $C_X$ denote the computational complexities of precoding optimization, power allocation optimization using Algorithm 1, and PA placement optimization using Algorithm 2, respectively, and $I_{\mathrm{AO}}$ denotes the number of AO iterations.
More specifically, the complexities of the precoding and power allocation subproblems are mainly dominated by matrix operations, while the complexity of the PA positions optimization is dominated by the repeated fitness evaluations in the PSO algorithm involving multiple particles and iterations. Therefore, when the number of PAs or the dimensionality of the search space increases, the computational cost is primarily dominated by the PSO module. Nevertheless, due to the inherent parallelism of PSO, the proposed method still maintains good scalability in practical implementations.

\section{Numerical Investigations}

We now evaluate the proposed algorithms through numerical experiments. Through the simulations, we analyze the effects of the BS transmit power, the number of PAs, the number of users, the side length of the user distribution area, and the number of waveguides on the WSR.

\subsection{Experimental Setting}

\begin{figure}
 \centering
  \includegraphics[width=7.6cm,height=6.1cm]{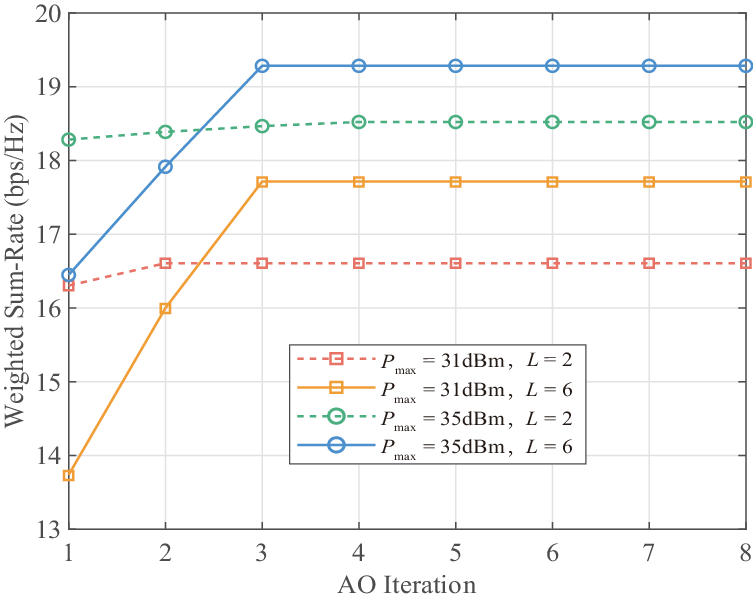}
  \caption{Convergence behavior of the AO iterative algorithm under different transmit powers $P_{\max}$ and PA numbers $L$. ($M=4$, $Z=6$, and $D=30$ m).}
  \vspace{-0.7cm}
  \label{fig:2}
  \end{figure}

\begin{figure}
 \centering
  \includegraphics[width=7.6cm,height=6.1cm]{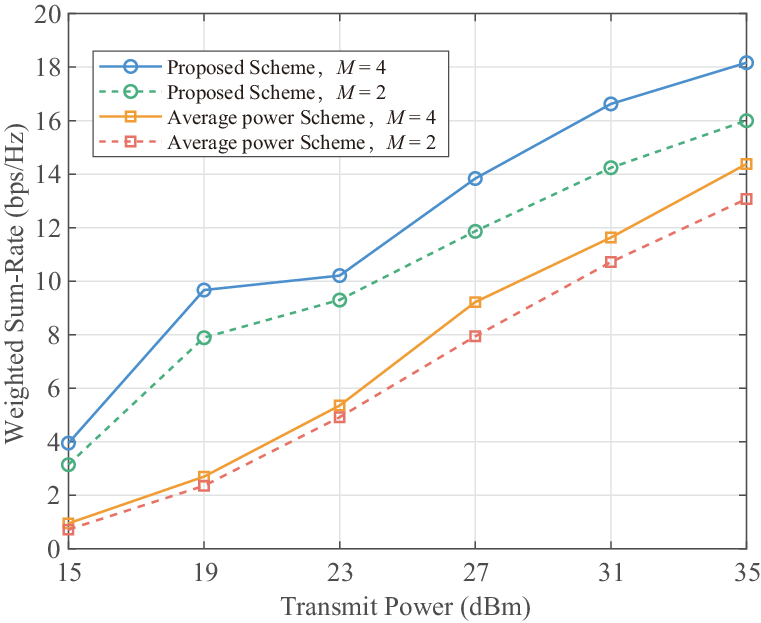}
  \caption{WSR versus transmit power $P_{\max}$ for different numbers of users $M$: comparison with the average power scheme. ($Z=6$, $L=6$, and $D=50$ m).}
  \vspace{-0.7cm}
  \label{fig:3}
  \end{figure}

We consider a PA-based  multi-cell communication system where the cell centers are uniformly distributed along the $x$-axis. Users are located on the $xy$-plane and are randomly distributed within a square region centered at each cell center with side length $D$. To emulate a more challenging scenario, most users are assumed to be edge users, i.e., they are randomly distributed closer to the boundaries of the region.

The system consists of $K=3$ cells, with centers at
$\left[-\frac{D}{2},0,0\right]$, $\left[0,0,0\right]$, and $\left[\frac{D}{2},0,0\right]$, as illustrated in Fig.~1. Each cell is equipped with $Z=6$ waveguides, and each waveguide is equipped with $L=6$ PAs to serve $M=4$ single-antenna users. The waveguides are assumed to be uniformly and symmetrically deployed above the service area at a height of $d=3$ m. The spacing between adjacent waveguides is set to $5$ m, and the waveguide length is equal to the side length $D$ of the considered region. The PAs are distributed along each waveguide. To avoid coupling effects, the minimum spacing between adjacent PAs on the same waveguide is set to $\Delta=\lambda/2$. The carrier frequency and noise power are set to $f_c=28$ GHz and $\sigma^2=-90$ dBm, respectively. The waveguide wavelength is given by $\lambda_g=\lambda/n_{\text{eff}}$, where $n_{\text{eff}}=1.44$ \cite{7}. Uniform user weights are adopted, i.e., $\alpha_{k,m}=1$ for $k \in \mathcal{K}$ and $m \in \mathcal{M}$.

During the optimization process, the convergence threshold of the proposed algorithm is set to $10^{-4}$. To ensure statistical reliability, all results are averaged over multiple Monte Carlo trials. Furthermore, when optimizing the PA positions using the PSO algorithm, multiple random restarts are performed in each simulation environment to reduce the risk of converging to local optima.

To evaluate the effectiveness of the proposed scheme and optimization method, we compare it with the following four baseline schemes.

\begin{itemize}

\item \textbf{Average Power Scheme}: In each cell, the BS transmit power is evenly allocated to all PAs on each waveguide, with no power allocation optimization performed.

\item \textbf{Fixed PA Placement}: In each cell, the PAs on each waveguide are uniformly distributed and their positions remain fixed throughout the entire optimization process.

\item \textbf{Conventional MIMO}: The system is equipped with $Z$ antennas deployed above the service area at height $d$, with an inter-element spacing of $\lambda/2$. Each antenna is connected to a dedicated radio-frequency (RF) chain, and full-digital signal processing is adopted. Specifically, the digital precoding is designed using the FP-BCD method, ensuring consistency with the precoder design adopted in the PA-based system.

\item \textbf{Massive MIMO}: The system employs an antenna array consisting of $Z\times L$ antennas deployed above the service area at height $d$. The antennas are arranged as a uniform planar array with an inter-element spacing of $\lambda/2$ along both the $x$- and $y$-axes. Each antenna is connected to a dedicated RF chain, and full-digital signal processing is adopted. Precoding is also designed using the FP-BCD method.

\end{itemize}

All schemes are evaluated under the same simulation settings for a fair comparison.

\subsection{Convergence}

\begin{figure}
 \centering
  \includegraphics[width=7.6cm,height=6.1cm]{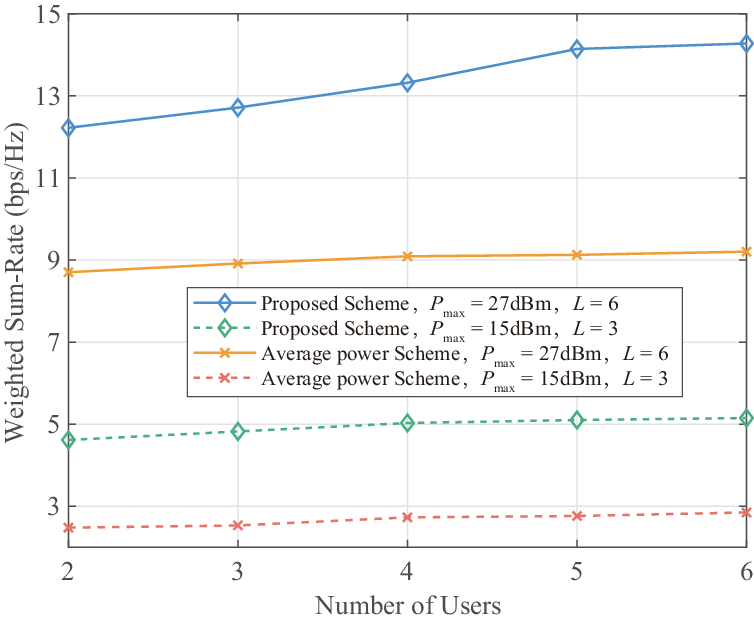}
  \caption{WSR versus the number of users $M$ under different transmit powers $P_{\max}$ and PA numbers $L$: comparison with the average power scheme. ($Z=6$ and $D=50$ m).}
  \vspace{-0.4cm}
  \label{fig:4}
  \end{figure}

\begin{figure}
 \centering
  \includegraphics[width=7.6cm,height=6.1cm]{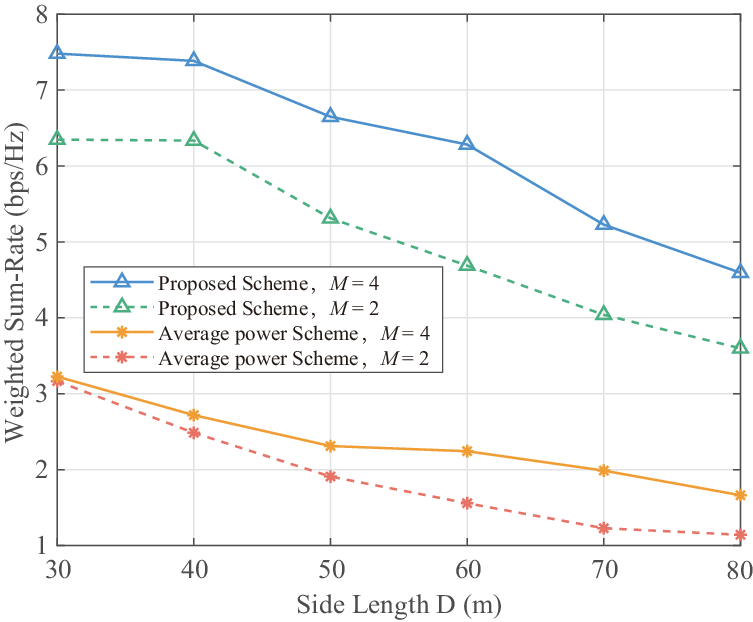}
  \caption{WSR versus the side length $D$ for different numbers of users $M$: comparison with the average power scheme. ($P_{\max}=19$ dBm, $Z=6$, and $L=6$).}
  \vspace{-0.7cm}
  \label{fig:5}
  \end{figure}
Fig.~2 illustrates the convergence behavior of the proposed algorithm under different BS transmit power levels and different numbers of PAs. Specifically, it can be observed that, for all considered parameter settings, the WSR increases monotonically with the number of iterations and converges to a stable value after only a few iterations, demonstrating the effectiveness and favorable convergence performance of the proposed algorithm.
In addition, both the BS transmit power and the number of PAs have significant impacts on the system performance. When the transmit power increases from $31$ dBm to $35$ dBm, the WSR improves significantly. Meanwhile, increasing the number of PAs on each waveguide can further enhance the system performance. This is because a larger number of PAs provides higher spatial DoFs, thereby improving signal enhancement and interference suppression capability.

\begin{figure}
 \centering
  \includegraphics[width=7.6cm,height=6.1cm]{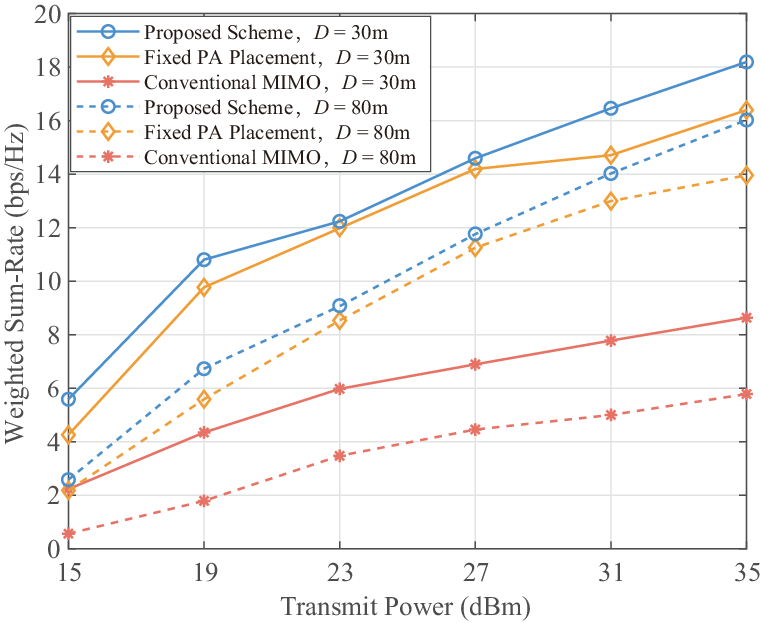}
  \caption{WSR versus transmit power $P_{\max}$ for different schemes under different side length $D$: comparison among the proposed scheme, fixed PA placement, and conventional MIMO. ($M = 4$, $Z = 6$, and $L = 6$).}
  \vspace{-0.4cm}
  \label{fig:6}
  \end{figure}

\begin{figure}
 \centering
  \includegraphics[width=7.6cm,height=6.1cm]{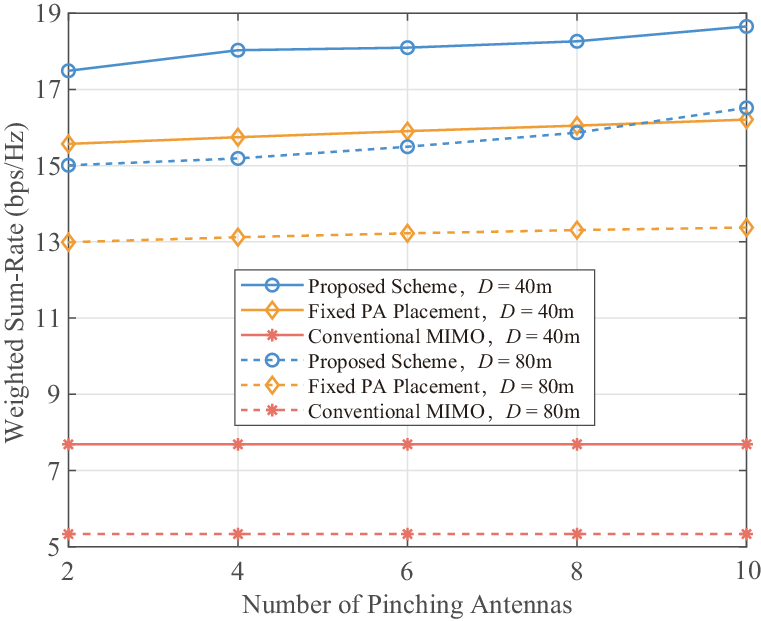}
  \caption{WSR versus the number of PAs $L$ for different schemes under different side length $D$: comparison among the proposed scheme, fixed PA placement, and conventional MIMO. ($M = 4$, $P_{\max}=35$ dBm, and $Z = 6$).}
  \vspace{-0.7cm}
  \label{fig:7}
  \end{figure}

\subsection{Impact of Power Allocation Optimization}

Figs.~3--5 illustrate the performance comparison between the proposed scheme and the average power allocation scheme under different system parameters.

Fig.~3 shows the WSR versus the maximum BS transmit power $P_{\max}$. It can be seen that the WSR increases monotonically with the transmit power. This is because a higher transmit power enhances the received signal strength and thus improves the SINR. Additionally, increasing the number of users further improves the WSR due to the enhanced spatial multiplexing gain.
Under the considered scenarios, the proposed scheme consistently outperforms the average power scheme. This performance gain is attributed to the optimization of the power allocation variable $\mathbf{P}_{k,z}$.

Specifically, under the constraint that the total transmit power of all PAs in each cell does not exceed the BS transmit power, the proposed scheme adaptively allocates power across different PAs according to the channel conditions of their associated transmission links. As a result, more power is assigned to links that contribute more to the system performance, i.e., links with favorable channel conditions or lower interference, thereby enhancing the received signal quality of users and improving the overall spectral efficiency.
This further demonstrates the necessity of PA power allocation in improving system performance under multi-user interference environments.

Fig.~4 illustrates the WSR versus the number of users $M$. As the number of users increases, the WSR gradually improves, although the growth rate becomes slower. This trend is due to the trade-off between spatial multiplexing gain and inter-user interference. Moreover, a higher transmit power and a larger number of PAs further enhance the system performance, as they provide more DoFs for signal enhancement and interference suppression.

Fig.~5 depicts the WSR versus the side length of the coverage area $D$. It can be observed that the WSR decreases as $D$ increases. This is mainly because a larger coverage area leads to a longer average transmission distance, resulting in more severe path loss. In addition, for each value of $D$, user locations are randomly generated, which introduces additional channel variations and slightly reduces the averaged WSR. Therefore, under the same system parameters, the WSR values in Fig.~5 are generally lower than those in Fig.~3 and Fig.~4, where the coverage area is fixed and the propagation conditions are more stable.

\subsection{Performance Impact of PA Placement Optimization: Comparison with Conventional MIMO}

Building upon the above results, Figs.~6--8 compare the performance of the proposed scheme, the fixed PA placement scheme, and the conventional MIMO scheme. It is observed that the proposed scheme consistently achieves the highest WSR across all considered scenarios. Compared with the fixed PA placement scheme, PA position optimization enables more accurate beam alignment and more effective interference suppression, which significantly reduces the path loss from the BS to the users. By contrast, the conventional MIMO scheme suffers from limited beamforming flexibility due to its fixed antenna structure, which restricts its ability to exploit spatial diversity, resulting in inferior performance.

In Fig.~6, the WSR of all schemes increases monotonically with the BS transmit power $P_{\max}$. Meanwhile, when the coverage size increases from $D=30$ m to $D=80$ m, the WSR decreases because of the longer propagation distance and more severe path loss, which is consistent with the observations in Fig.~5.

\begin{figure}
 \centering
  \includegraphics[width=7.6cm,height=6.1cm]{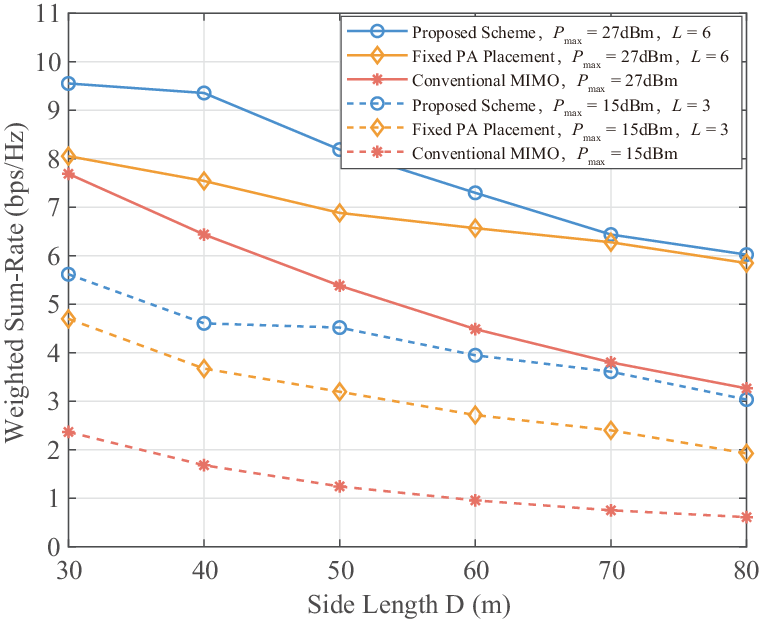}
  \caption{WSR versus the side length $D$ for different schemes under different transmit powers $P_{\max}$ and PA numbers $L$: comparison among the proposed scheme, fixed PA placement, and conventional MIMO. ($M = 4$ and $Z = 6$).}
  \vspace{-0.4cm}
  \label{fig:8}
  \end{figure}

\begin{figure}
 \centering
  \includegraphics[width=7.6cm,height=6.1cm]{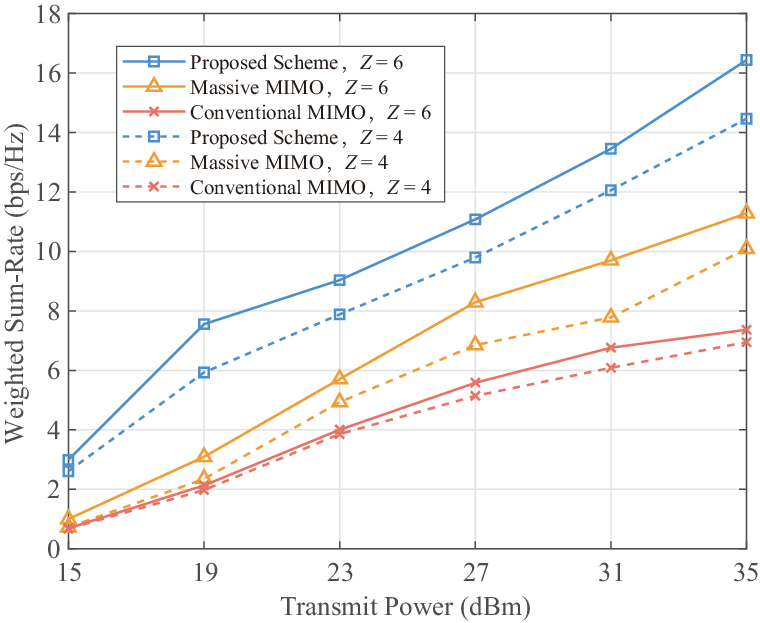}
  \caption{WSR versus transmit power $P_{\max}$ for different schemes under different numbers of waveguides $Z$: comparison among the proposed scheme, massive MIMO, and conventional MIMO. ($M = 4$, $L = 6$, and $D=70$ m).}
  \vspace{-0.65cm}
  \label{fig:9}
  \end{figure}

Fig.~7 illustrates the WSR versus the number of PAs, where the horizontal axis represents the number of PAs deployed on each waveguide. For both the proposed scheme and the fixed PA placement scheme, the WSR increases with the number of PAs. This is because more PAs provide higher spatial DoFs and array gain, thereby enhancing the beamforming capability.
However, since the proposed scheme jointly optimizes the PA positions, precoding matrix, and power allocation, it can efficiently exploit the available spatial DoFs even with a relatively small number of PAs, achieving accurate beamforming and effective energy focusing. Meanwhile, since the total number of PAs always exceeds the number of users, the spatial DoFs gradually become saturated as the number of PAs increases. In addition, the system performance is constrained by multiuser interference and total power budget, which leads to diminishing marginal gains from additional PAs and results in a slower growth of the WSR.
For the fixed PA placement scheme, newly added PAs cannot be adaptively adjusted according to the user distribution, leading to inefficient utilization of spatial resources. As a result, the performance improvement is also limited. In contrast, the number of antennas in the conventional MIMO scheme is solely determined by the number of waveguides $Z$, and thus it lacks additional spatial flexibility to benefit from increasing the number of PAs, resulting in the worst performance among all schemes.

Fig.~8 shows the WSR versus the coverage size $D$ under different transmit powers $P_{\max}$ and PA numbers $L$. It can be seen that a higher transmit power consistently leads to a higher WSR for all schemes, indicating that transmit power remains a dominant factor affecting system performance. The reason why the WSR decreases with increasing $D$ has already been discussed in Fig.~5 and is thus omitted here for brevity. Nevertheless, the proposed scheme still maintains a clear performance advantage across all coverage sizes, demonstrating its robustness in large-scale propagation environments.

\subsection{Comparison with Massive MIMO Architectures}
In the previous subsection, conventional MIMO was introduced as a benchmark scheme, whose performance is limited by its fixed antenna structure and relatively small number of antennas. Therefore, in this subsection, massive MIMO is further considered as a more competitive baseline to provide a more comprehensive evaluation of the proposed scheme. Figs.~9--11 illustrate the WSR performance comparison among the proposed scheme, massive MIMO, and conventional MIMO under different system parameters.

\begin{figure}
 \centering
  \includegraphics[width=7.6cm,height=6.1cm]{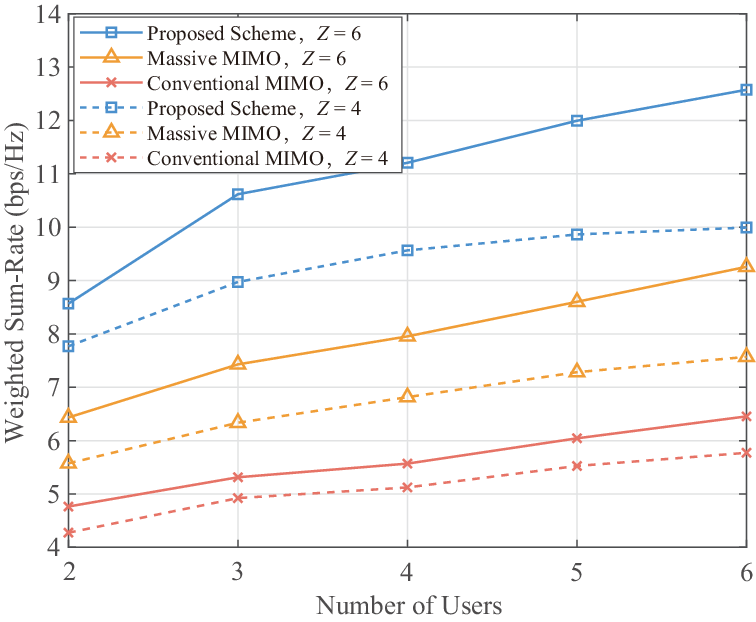}
  \caption{WSR versus the number of users $M$ for different schemes under different numbers of waveguides $Z$: comparison among the proposed scheme, massive MIMO, and conventional MIMO. ($P_{\max}=27$ dBm, $L = 6$, and $D=70$ m).}
  \vspace{-0.7cm}
  \label{fig:10}
  \end{figure}
\begin{figure}
 \centering
  \includegraphics[width=7.6cm,height=6.1cm]{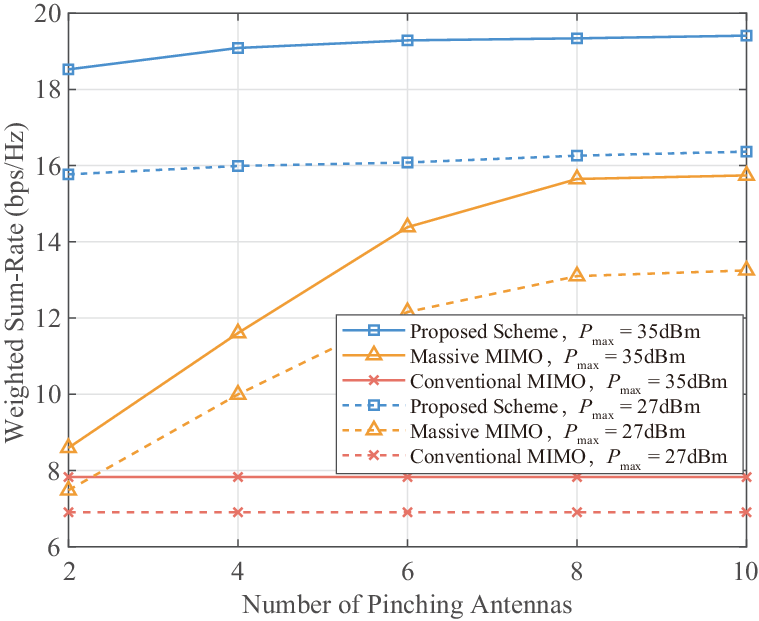}
  \caption{WSR versus the number of PAs $L$ for different schemes under different transmit powers $P_{\max}$: comparison among the proposed scheme, massive MIMO, and conventional MIMO. ($M = 4$, $Z = 6$, and $D=30$ m).}
  \vspace{-0.7cm}
  \label{fig:11}
  \end{figure}

Specifically, Fig.~9 shows the WSR versus the transmit power $P_{\max}$. It can be observed that the WSR of all schemes increases significantly with the number of waveguides $Z$. This is because a larger $Z$ corresponds to more RF chains and spatial DoFs, thereby enhancing the signal transmission capability and spatial multiplexing gain. The proposed scheme consistently outperforms both massive MIMO and conventional MIMO, since it enables more flexible spatial reconfiguration by optimizing the PA positions, resulting in higher array gain and more effective multiuser interference suppression.

Fig.~10 presents the WSR versus the number of users $M$. As the number of users increases, the WSR continuously improves. Moreover, increasing the number of waveguides $Z$ further enhances the system performance. In contrast, the performance improvement of conventional MIMO is relatively limited, since its fixed antenna structure cannot efficiently exploit the additional spatial resources. It is worth noting that when $M \geq 4$, the performance gap between the cases of $Z=6$ and $Z=4$ becomes more pronounced for the proposed scheme. This is because when $M \geq Z$, the system gradually loses the ability to provide sufficient spatia DoFs to support all data streams, and the interference between users gradually increases.

Fig.~11 illustrates the WSR versus the number of PAs. The proposed scheme achieves the highest WSR across all PA numbers. Meanwhile, the WSR of massive MIMO increases significantly with the number of PAs, since its performance mainly relies on enlarging the antenna array to obtain higher array gain. When the number of PAs is small, its performance is limited, but it gradually improves as more PAs are deployed. However, even with a large number of PAs, massive MIMO still underperforms the proposed scheme, indicating that merely increasing the antenna scale cannot replace the performance gains brought by spatial reconfiguration via PA position optimization.

\section{Conclusion}

To address the optimization challenges arising from the strong coupling among precoding, power allocation, and antenna placement in PA-assisted multi-cell communication systems, this paper developed an AO-based joint design framework to maximize the WSR.
Specifically, FP is first employed to reformulate the original nonconvex problem into an equivalent form, where auxiliary variables are introduced to effectively decouple the signal and interference terms. Based on this reformulation, a BCD method is adopted to iteratively update the optimization variables, where both the precoding matrices and power allocation variables can be efficiently obtained via analytical expressions. For the high-dimensional and nonconvex antenna position subproblem, a PSO algorithm is utilized to perform efficient search. By iteratively updating the positions and velocities of particles, the algorithm gradually approaches the optimal PA configuration, thereby improving solution quality and enhancing scalability.
The results indicate that jointly optimizing the spatial domain and signal domain effectively enhances the spatial DoFs, thereby significantly improving system performance in multiuser scenarios.
Numerical results further validate the superiority of such joint optimization strategies over conventional antenna systems in complex multi-cell environments, demonstrating the significant potential of PA technology in mitigating inter-cell interference and enhancing system capacity. These findings reveal the critical role of flexible antenna architectures in future wireless systems, particularly for large-scale and dense deployments.

Future work may extend this framework to more complex channel environments and practical system constraints,
such as frequency selective fading channels and imperfect channel state information,
facilitating the deployment of PA technologies in real-world wireless networks.

\ifCLASSOPTIONcaptionsoff
  \newpage
\fi

\bibliographystyle{IEEEtran}
\bibliography{reference}

\end{document}